\documentclass[aps,prd,twocolumn,floatfix,nofootinbib,showpacs,superscriptaddress]{revtex4-1}

\usepackage{amsmath,amsfonts,amssymb,bm}
\usepackage{epsfig}
\usepackage{graphicx}
\usepackage{color}
\usepackage{pzccal}
\usepackage{soul}

\newcommand{\be}{\begin{eqnarray}}
\newcommand{\ee}{\end{eqnarray}}

\newcommand{\bea}{\begin{eqnarray}}
\newcommand{\eea}{\end{eqnarray}}
\newcommand{\beas}{\begin{eqnarray*}}
\newcommand{\eeas}{\end{eqnarray*}}

\begin{document}
\title{Multiple solutions for the fermion mass function in QED3}
\author{K. Raya}
\affiliation{Instituto de F\1sica y Matem\'aticas, Universidad
Michoacana de San Nicol\'as de Hidalgo,
Edificio C-3, Ciudad Universitaria, C.P. 58040,
Morelia, Michoac\'an, Mexico}

\author{A. Bashir}
\affiliation{Instituto de F\1sica y Matem\'aticas, Universidad
Michoacana de San Nicol\'as de Hidalgo,
Edificio C-3, Ciudad Universitaria, C.P. 58040,
Morelia, Michoac\'an, Mexico}

\author{S. Hern\'andez-Ortiz}
\affiliation{Instituto de F\1sica y Matem\'aticas, Universidad
Michoacana de San Nicol\'as de Hidalgo,
Edificio C-3, Ciudad Universitaria, C.P. 58040,
Morelia, Michoac\'an, Mexico}

\author{A. Raya}
\affiliation{Instituto de F\1sica y Matem\'aticas, Universidad
Michoacana de San Nicol\'as de Hidalgo,
Edificio C-3, Ciudad Universitaria, C.P. 58040,
Morelia, Michoac\'an, Mexico}

\author{C. D. Roberts}
\affiliation{Physics Division, Argonne National Laboratory, Argonne, Illinois 60439, USA}
\affiliation{Department of Physics, Illinois Institute of
Technology, Chicago, Illinois 60616-3793, USA}

\date{15 July 2013}

\begin{abstract}

\medskip

\centerline{\parbox[c]{0.6\linewidth}{\emph{This work is dedicated to the memory of Alfredo Raya Sr., guiding father, mentor and friend.  Your legacy will live on in those whose lives you touched.  Descanse en Paz.}}}

\bigskip

Theories that support dynamical generation of a fermion mass gap are of widespread interest. The phenomenon is often studied via the Dyson-Schwinger equation (DSE) for the fermion self energy; i.e., the gap equation.  When the rainbow truncation of that equation supports dynamical mass generation, it typically also possesses a countable infinity of simultaneous solutions for the dressed-fermion mass function, solutions which may be ordered by the number of zeros they exhibit.  These features can be understood via the theory of nonlinear Hammerstein integral equations.  Using QED3 as an example, we demonstrate the existence of a large class of gap equation truncations that possess solutions with damped oscillations.  We suggest that there is a larger class, quite probably including the exact theory, which does not.  The structure of the dressed-fermion--gauge-boson vertex is an important factor in deciding the issue.
%
\end{abstract}

\maketitle

\section{Introduction}\label{secI}
Dynamical mass generation in gauge theories has long been studied using the gap equation; i.e., the Dyson-Schwinger equation (DSE) for the dressed fermion propagator \cite{Roberts:1994dr,Bashir:2012fs}.  The gap equation is nonlinear in the dressed-fermion mass-function, $M(p)$; and it was found early \cite{Fukuda:1976zb} that a simple (rainbow) truncation of this equation possess a countable infinity of solutions, with each one characterized by: its magnitude and sign at $p=0$; and the number of zeros it possesses on $p\in (0,\infty)$.  The existence of this tower of solutions to the truncated equation almost immediately fostered speculation about their possible physical consequences \cite{Miransky:1980vc,Miransky:1981rt,Fomin:1984tv,Miransky:1999tr,Gorbar2003472}.

The non-uniqueness of solutions to the rainbow gap equation was later rediscovered independently, with the explicit identification of three distinct solutions that bifurcate from the trivial $M(p)\equiv 0$ solution as the coupling strength is increased \cite{Cheng:1994sd}; and then again about a decade later \cite{Chang:2006bm,Martin:2006qd,Williams:2006vva}.  More recently, solutions of the gap equation with repeated zero crossings were also found using a simple generalization of the rainbow truncation \cite{Fischer:2008sp}; and in Ref.\,\cite{Wang:2012me} their existence was firmly established using a procedure that enables one to obtain all solutions of the gap equation in the presence of a variety of external control parameters.

In considering all these contributions one is naturally led to wonder about the general conditions under which a gauge theory's gap equation will admit solutions with zeros in addition to that single positive-definite solution which is conventionally associated with dynamical chiral symmetry breaking (DCSB).  Herein we provide a partial answer, using quantum electrodynamics in three dimensions (QED3) as an illustrative tool.

Our choice of theory is motivated first by a number of properties that QED3 shares with quantum chromodynamics (QCD).  For example, quenched QED3 possesses a nonzero string tension \cite{Gopfert:1981er}; and this feature persists in the unquenched theory if massive fermions circulate in the photon vacuum polarization \cite{Burden:1991uh}.\footnote{This persistence contrasts with unquenched QCD, however, which does not possess a measurable string tension in the neighborhood of light quarks \cite{Bali:2005fu}.}  In addition, since QED3 is super-renormalizable, it has a well-defined chiral limit and therefore admits the possibility and study of DCSB.  DCSB in QCD underlies the success of chiral effective field theory, explains the origin of constituent-quark masses \cite{Bhagwat:2003vw,Bhagwat:2006tu,Roberts:2007ji} and hence the vast bulk of visible mass in the Universe \cite{national2012Nuclear}, and quite probably shares its origin with light-quark confinement \cite{Roberts:2011ea}.  There is also an applied interest and relevance because QED3 is used in condensed matter physics as an effective field theory for high-temperature superconductors \cite{Franz:2002qy,Herbut:2002yq,Thomas:2006bj} and graphene \cite{Novoselov:2005kj,Gusynin:2007ix,Gamayun:2009em}.

Our partial answer to the problem of anticipating the appearance of solutions to the gap equation with one or more damped oscillations is based upon the following observation \cite{Pagels:1979ai}: in rainbow truncation the dressed-fermion mass function satisfies a nonlinear Hammerstein integral equation \cite{Polyanin:2008}.  The conditions under which such equations admit oscillatory solutions were long ago elucidated in the mathematics literature \cite{Pimbley,Parter,Crandall}.  We find that if the dressed-fermion-photon vertex produces a gap equation kernel that satisfies those conditions, then solutions with zeros are readily found numerically.  This is not the case when the vertex produces a kernel that violates the conditions.

Our material is organized as follows.  In Sec.\,\ref{Theory} we recapitulate briefly upon the theory of nonlinear Hammerstein integral equations, listing the criteria which guarantee the existence of oscillatory solutions for the mass function and providing a simple example.  We analyze QED3 from the perspective of Hammerstein integral equations in Sec.\,\ref{QED3-gap}; and in Sec.\,\ref{gauge} we explain that if oscillatory solutions exist in Landau gauge, then they are present in all covariant gauges.  We summarize in Sec.\,\ref{final}.

\section{Hammerstein integral equations}
\label{Theory}
\subsection{Definition}
\label{Theory-H}
A nonlinear, homogeneous Hammerstein integral equation of the first kind has the form \cite{Polyanin:2008}
\begin{equation}
M(x) = \lambda\int_0^1dy\, G(x,y)\, H\left(y,M(y)\right)M(y)\,,
\label{eq:Hammer}
\end{equation}
where $M(x)$ is the equation's solution, $H\left(x,M(x)\right)$ is a nonlinear functional of $M(x)$, $G(x,y)$ is the kernel of the equation and $\lambda$ is a real number.  A solution of Eq.\,(\ref{eq:Hammer}) is a pair $(\lambda,M)$, where $\lambda\in\mathbb{R}$ and $M(x)$ is a continuous function on $x\in [0,1]$.  Suppose now that:
\begin{itemize}
\item[$\mathbf{H_1}$:] $H(y,z)$ and $H_z(y,z):=\partial_z H(y,z)$ are continuous, and $H(y,z)\!>\!0$ for $\{y,z\}$ $\in[0,1]\times\mathbb{R}$;

\item[$\mathbf{H_2}$:] $H(y,z)+zH_z(y,z)>0$ for $\{y,z\}\in[0,1]\times\mathbb{R}$;

\item[$\mathbf{H_3}$:] $zH_z(y,z)<0$ for $y\in[0,1]$ and $z\neq 0$;

\item[$\mathbf{H_4}$:] $H(y,z)\rightarrow 0$ when $|z|\rightarrow\infty $ uniformly for $y\in[0,1]$;

\item[$\mathbf{H_5}$:] $H(y,z)=H(y,-z)$ for $\{y,z\}\in[0,1]\times\mathbb{R}$; and

\item[$\mathbf{G_1}$:] $G(x,y)$ is a symmetric oscillation kernel\footnote{The nature of oscillation kernels is explained, e.g., in Ref.\,\protect\cite{Gantmakher}.  Put simply, they are a class of positive, symmetric kernels which, by nature, support oscillatory solutions.} on $\{x,y\}\in [0,1]\times[0,1]$.
\end{itemize}
In such circumstances it is known \cite{Pimbley,Parter} that if $\lambda$ exceeds
$\lambda_{j}$, where $\lambda_j$, $j\geq 1$, is the $j$-th eigenvalue of the linear operator associated with Eq.\,(\ref{eq:Hammer}); i.e., the integral equation obtained through the replacement
\begin{equation}
H(y,M(y))\to H(y,0)\,,
\end{equation}
then the Hammerstein equation possesses at least $2j$ nontrivial solutions with zeros in $(0,1)$.  It was subsequently demonstrated \cite{Crandall} that Eq.\,(\ref{eq:Hammer}) possesses infinitely many solutions, each distinguished by its number of zeros, for any prescribed value of
\begin{equation}
{\sup_{\substack{0\le x\le 1}}}|M(x)|.
\end{equation}
Thus, as with differential equations, at least one large class of integral equations possesses an enumerable infinity of solutions with damped oscillations.

\subsection{Example}
A simple, physical realization of the Hammerstein integral equation may arise in connection with graphene, a one-atom-thick layer of graphite, wherein quasiparticle excitations are described by the massless Dirac equation in $2+1$-dimensions.  This is the Euler-Lagrange equation for fermions in a $2+1$-dimensional version of quantum electrodynamics, except that the speed of light is replaced by the Fermi velocity; viz.,
\begin{equation}
c \approx 3\times 10^8 {\rm ms}^{-1} \to v_{F} \approx 1 \times 10^{6}{\rm ms}^{-1}.
\end{equation}
Thus, the interaction strength in suspended graphene is:
\begin{equation}
 \alpha = \frac{e^2}{\hbar c} \approx \frac{1}{137} \to
 \alpha_{\rm eff} = \frac{e^2}{\hbar v_{\rm eff}} \approx 2 \;,
\end{equation}
in which case the model's gap equation can produce solutions that dynamically break chiral symmetry; and such solutions express a realignment of the ground-state within the sample.  For graphene on a substrate, however, the effective coupling is screened, owing to a  dielectric constant, $\varepsilon$, associated with the substrate:
\begin{equation}
 \alpha_{\rm eff} \to \alpha_{\rm sub} = \frac{\alpha_{\rm eff}}{\varepsilon}\,;
\end{equation}
and for $\varepsilon$ sufficiently large, the symmetry breaking solutions will disappear.

\begin{figure}[t] 
\centerline{%
\includegraphics[clip,width=0.8\linewidth]{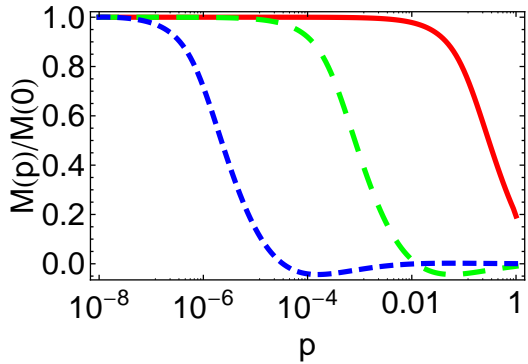}}
\caption{Solutions of Eq.\,(\ref{eqgraf}), plotted as $M(p)/M(0)$.  Solid curve -- no zeros; long-dashed curve -- one zero; and short-dashed curve -- two zeros.  With $\alpha = \pi/2$, in units of $\Lambda \simeq 1/a$, the true magnitude of the solutions is, respectively, O$(10^{-1})$, O$(10^{-4})$, O$(10^{-7})$. \label{fig:graphene}}
\end{figure}

The phase transition in this model is described by the following gap equation \cite{Gamayun:2009em}
\begin{equation}
M(x) =  \frac{\alpha}{\pi}\int_0^1 dy \,\mathcal{G}(x,y)
\frac{y}{\sqrt{y^2+M^2(y)}} \,M(y)\,, \;\label{eqgraf}
\end{equation}
with
\begin{equation}
\mathcal{G}(x,y)=\frac{1}{x+y}K\left(\frac{2\sqrt{x
y}}{x+y}\right),
\end{equation}
where $K(z)$ is the complete elliptic integral of the first kind and all mass-dimensioned quantities have been rescaled by an ultraviolet cutoff $\Lambda \simeq 1/a$, where $a$ characterizes the lattice spacing in the sample.   It is straightforward to verify that the elements in Eq.\,\eqref{eqgraf} satisfy the conditions $\mathbf{H_1}$--$\mathbf{H_5}$, $\mathbf{G_1}$ and hence there is an $\alpha_c$ such that for $\alpha>\alpha_c$ the gap equation possesses a countable infinity of distinct solutions, distinguished one from another by the number of zeros they exhibit.  In this case, $\alpha_c = 1/2$ \cite{Gamayun:2009em}; and the first three solutions are depicted in Fig.\,\ref{fig:graphene}.

\section{QED3 Gap Equation}
\label{QED3-gap}
\subsection{Problem specification}
The fully dressed fermion propagator is commonly written in one of the following forms:
\begin{subequations}
\begin{eqnarray}
S(p)^{-1} &=& i \gamma\cdot p + m_0 + \Sigma(p) \,,\\
&=& Z(p)/ [ i\gamma\cdot p  + M(p) ]\,,\\
&=& -i \gamma\cdot p\, \sigma_V(p) + \sigma_S(p)\,, \label{Spsigmavs}
\end{eqnarray}
\end{subequations}
where $m_0$ is a current-fermion Lagrangian mass and the self-energy is given by
\begin{eqnarray}
\Sigma(p) &=& e^2  \int \frac{d^3 k}{(2\pi)^3}  D_{\mu\nu}(p-k) \gamma_\mu S(k) \Gamma_\nu(k,p) \,.
\label{Self-energy}
\end{eqnarray}
Here, $D_{\mu\nu}(p-k)$ and $\Gamma_\nu(k,p)$ are the dressed photon propagator and dressed
fermion-photon vertex, respectively; and $e^2$ is the coupling, which has mass dimension one.  Since QED3 is super-renormalizable and therefore no ultraviolet divergence can arise whose regularization would introduce a new mass-scale, $e^2$ defines the natural scale of the theory.  

The quenched, rainbow truncation of Eq.\,\eqref{Self-energy} is obtained with
\begin{eqnarray}
D_{\mu\nu}(q) &\to& \left[\delta_{\mu\nu} - \frac{q_\mu q_\nu}{q^2}\right]\frac{1}{q^2}\\
\Gamma_\nu(q,p) & \to & \gamma_\nu\,.
\end{eqnarray}
We work primarily in Landau gauge because it occupies a special place in gauge theories \cite{Bashir:2008fk,Bashir:2009fv}.  It is the gauge in which any sound \emph{Ansatz} for the fermion-photon vertex can most legitimately be described as providing a pointwise accurate approximation.  The vertex in any other gauge is then defined as the Landau-Khalatnikov- Fradkin (LKF) transform \cite{Fradkin:1955jr,LK56,Johnson:1959zz,Zumino:1959wt} of the Landau gauge \emph{Ansatz}.  The sensible implementation of this procedure guarantees gauge covariance and hence obviates any question about the gauge dependence of gauge invariant quantities.  We expand on these points in Sec.\,\ref{gauge}.

After a little Dirac algebra and evaluation of the angular integrals in Eq.\,\eqref{Self-energy}, one finds $Z(p)\equiv 1$ and a single equation for the dressed-fermion mass function ($\alpha = e^2/[4\pi]$):
\begin{equation}
M(p) = m_0+ \frac{2 \alpha}{\pi p} \int_{0}^{\infty}\! dk \, \frac{k
M(k)}{k^2 + M^2 (k)} \ln  \left| \frac{k+p}{k-p}\right|\;.
\label{SDE-Mass}
\end{equation}

One may now exploit the fact that QED3 is a super-renormalizable theory, in which the explicit mass-scale is defined by $e^2$ and effects associated with dynamical mass generation are an order of magnitude smaller (see Fig.\,\ref{fig:graphene}).  Ultraviolet momenta therefore have no influence on nonperturbative phenomena and hence, at no cost, one may introduce a mass-scale $\Lambda \gg m_0$, $\Lambda \gg e^2$ but $m_0/\Lambda$, $e^2/\Lambda$ fixed, such that the following gap equation is equivalent to Eq.\,\eqref{SDE-Mass}:
\begin{equation}
\tilde M(x) = \tilde m_0+ \frac{2 \tilde \alpha}{\pi x} \int_{0}^{1}\! dy \, \frac{y
\tilde M(y)}{y^2 + \tilde M^2 (y)} \ln  \left| \frac{y+x}{y-x}\right|\;,
\label{SDE-Mass1}
\end{equation}
where $\tilde M(x) = M(p/\Lambda)/\Lambda$, $\tilde m_0 = m_0/\Lambda$, $\tilde \alpha = \alpha/\Lambda$.

In some of our subsequent analysis and illustrations, we will employ the approximation
\begin{equation}
\ln \left| \frac{y+x}{y-x} \right| \simeq \frac{2x}{y}\ \theta(y-x)+\frac{2y}{x}\
\theta(x-y)\;,\label{log}
\end{equation}
which is precise for $y\gg x$ and $x \ll y$, and, in fact, quite accurate in general for such a weakly singular kernel \cite{Roberts:1989mj}.  This leads from Eq.\,\eqref{SDE-Mass1} to
\begin{eqnarray}
\nonumber
\tilde M(x) &=& \tilde m_0+ \frac{4 \tilde \alpha}{\pi x} \int_{0}^{1}\! dy \, \frac{y
\tilde M(y)}{y^2 + \tilde M^2 (y)} \rule{8em}{0ex}\\
&& \rule{5em}{0ex}\times \bigg[
\frac{x}{y}\ \theta(y-x)+\frac{y}{x}\theta(x-y)
\bigg]
\;,
\label{SDE-MassLog}
\end{eqnarray}
an inhomogeneous Hammerstein equation, which becomes homogeneous when $\tilde m_0 = 0$.

\subsection{Linearized gap equation}
\label{sec:linearized}
If one sets $\tilde M^2(y)\equiv 0$ in the denominator of the integrand in Eq.\,\eqref{SDE-MassLog}, then one arrives at a linear equation that can be treated analytically.  In the chiral limit, however, unlike its parent, that equation is invariant under rescaling: $\tilde M(x) \to c \tilde M(x)$, with $c$ an arbitrary constant, and does not possess an infrared regularizing mechanism.  These qualitative differences are both remedied by the simple expedient of introducing an infrared cutoff, $\kappa$, with $\tilde M(x=\kappa)=\kappa$:
\begin{equation}
\tilde M(x) = \frac{4 \tilde \alpha}{\pi x} \int_{\kappa}^{1}\! dy \,
\frac{\tilde M(y)}{y}
\bigg[
\frac{x}{y}\ \theta(y-x)+\frac{y}{x}\theta(x-y)
\bigg]
\;.
\label{SDE-MassLogK}
\end{equation}
If one relaxes $\mathbf{H_2}$ mildly, to the extent that the condition is satisfied on $[\kappa,1]\times\mathbb{R}$, this is an homogeneous Hammerstein equation.

\begin{figure}[t] 
\centerline{%
\includegraphics[clip,width=0.8\linewidth]{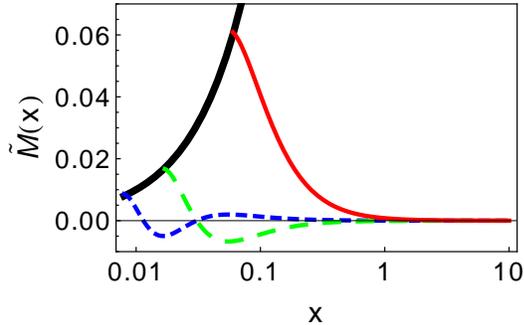}}
\caption{Solutions of Eq.\,\protect\eqref{SDE-MassLogK}; viz., linearized rainbow-ladder gap equation for quenched QED3.  \emph{Solid curve} is $\hat M(x) = \kappa$.  The nature of the solution evolves with $\kappa$: it acquires an additional zero each time $\kappa$ decreases through one of a countable infinity of thresholds.  \label{fig:nodan}}
\end{figure}

So long as the following boundary conditions are applied:
\begin{equation}
\tilde M(\kappa) = \kappa\,,\;
\tilde M(x)\Big|_{x\to\infty}=0\,,
\end{equation}
Eq.\,\eqref{SDE-MassLogK} is equivalent to the following second-order differential equation:
\begin{equation}
x^3 \tilde M^{\prime\prime}(x)+3x^2\tilde M^\prime(x)+\frac{\mathcal a}{2}\tilde M(x)=0\,,
\end{equation}
where $\mathcal a=16 \tilde \alpha/\pi$.  This problem has the solution
\begin{eqnarray}
\tilde M(x)&=& \mathcal n \frac{\mathcal a}{x}J_2\left(\sqrt{2 \mathcal a/x}\right)\,, \\
\mathcal n &=& \frac{\kappa^2}{\mathcal a} \frac{1}{J_2(\sqrt{2 \mathcal a/\kappa})}\,,
\end{eqnarray}
where $J_2(z)$ is a Bessel function.  For $\kappa \sim \mathcal a$, the solution is monotonically decreasing and positive definite.  This remains true as $\kappa$ is reduced until, at $\kappa \simeq \mathcal a/14$, there is a qualitative change in the solution, with the appearance of a zero.  A second zero appears for $\kappa \simeq \mathcal a/36$; and this pattern, illustrated in Fig.\,\ref{fig:nodan}, continues \emph{ad infinitum}.

This discussion illustrates the statements of Sec.\,\ref{Theory-H} and may be viewed as establishing that, in the chiral limit, the gap integral equation, Eq.\,\eqref{SDE-Mass1}, possesses a countably infinite number of simultaneous solutions, which are distinguished by their magnitude and the correlated number of zeros.

Sturm-Liouville theory provides another perspective on the guaranteed simultaneous existence of a countable infinity of solutions to Eq.\,\eqref{SDE-Mass1}.  Consider the linearized version of Eq.\,\eqref{SDE-Mass1} in the chiral limit:
\begin{equation}
\tilde M(x) = \frac{2 \tilde\alpha}{\pi} \int_{\kappa}^{1}dy \,
\frac{\tilde M(y)}{y x} \ln \left| \frac{y+x}{y-x}\right|\,. \label{linearizedI}
\end{equation}
The solutions of this equation cannot be obtained analytically.  However, it can be viewed as one in a class of integral equations with the general form:
\begin{equation}
\lambda f(x) = \int_a^b dy\, f(y)\, G_f(y,x)   \,,
\end{equation}
where the kernel $G_f(y,x)$ is real and symmetric under $x\leftrightarrow y$.  This is an homogeneous Fredholm equation of the first kind.

Introducing a $n$-point quadrature rule, Eq.\,\eqref{linearizedI} is replaced by a system of coupled algebraic equations:
\begin{equation}
\label{QuadratureI}
\lambda \tilde M_i = \sum_{j=1}^n K_{ij} A_j  \tilde M_j,
\end{equation}
where $\tilde M_i=\tilde M(x_i)$, $\tilde M_j=\tilde M(y_j)$, and the set $\{A_j>0,j=1,\ldots,n\}$ contains the weights associated with the chosen quadrature.  One may always choose a rule with $p_i=k_i, \forall i$, so that $K_{ij}$ corresponds to a symmetric matrix.  Then, defining $\Psi_i = A_i^{1/2} \tilde M_i$ and $S_{ij}=A_i^{1/2} K_{ij} A_j^{1/2}$, Eq.\,\eqref{QuadratureI} can be expressed
\begin{equation}
\lambda \Psi = S \Psi\,. \label{Liouville}
\end{equation}
This is an eigenvalue problem for a real, symmetric matrix, $S$; and different eigenvectors correspond to distinct solutions of the gap equation.

\begin{figure}[t] 
\centerline{%
\includegraphics[clip,width=0.8\linewidth]{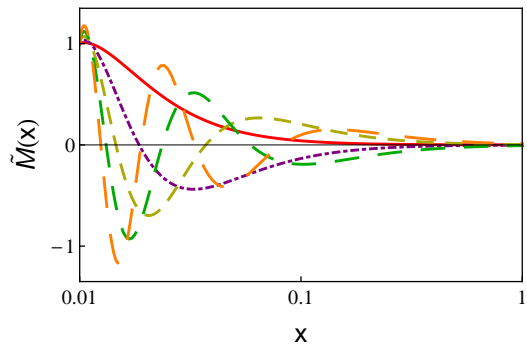}}
\caption{Solutions of Eq.\,\eqref{linearizedI} associated with the four smallest eigenvalues: solid curve, no zeros; long-dashed curve, one zero; dashed curve, two zeros; and dot-dashed curve, three zeros.  All curves have been rescaled such that $|\tilde M(0)=1|$. \label{fig:num}}
\end{figure}

We have applied the procedure just described to Eq.\,\eqref{linearizedI}, using a Gaussian quadrature.  The first four eigenfunctions are depicted in Fig.\,\ref{fig:num}.  All solutions fall as $1/x^2$ in the ultraviolet and all possess the Sturm-Liouville property; namely, the $n^{\rm th}$ eigenfunction, $\Psi^{(n)}$, possesses one more zero than $\Psi^{(n-1)}$.  The number of independent eigenfunctions equals the number of quadrature points and hence, in the continuum limit, there is a countable infinity of solutions, each distinguished by its magnitude and number of zeros.



\subsection{Nonlinear gap equation}
\label{sec:nonlinear}
\begin{figure}[t] 
\centerline{
\includegraphics[clip,width=0.8\columnwidth]{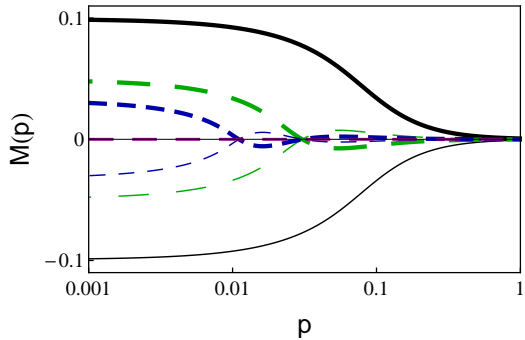}}
\caption{First three nontrivial solutions of Eq.\,\protect\eqref{SDE-Mass}: no zeros (solid curve); one zero (dashed curve); and two zeros (short-dashed curve).  The thin curves are their mirror solutions. (Mass-dimension is fixed by setting $e^2=1$.) \label{fig:allnoded}}
\end{figure}

Armed with the knowledge accumulated above, we are ready to find all solutions of the original problem; i.e., Eq.\,\eqref{SDE-Mass}.  In the chiral limit, besides the trivial $M(p)\equiv 0$ solution that is admitted in perturbation theory, and the well-known positive-definite DCSB solution, we find the now expected series of solutions, each one of which may be labeled uniquely by the number of zeros it possesses and its sign at $p=0$.  One must include that sign because the chiral-limit gap equation is even under $M\to -M$, so each solution has a mirror image.  The first few solutions are illustrated in Fig.\,\ref{fig:allnoded}.  All nonzero solutions exhibit at least one inflection point in the infrared\footnote{The existence of an inflection point can be understood as a signal of confinement \protect\cite{Roberts:2007ji,Bashir:2008fk,Bashir:2009fv,Roberts:2011ea}.} and decay as $1/p^2$ in the ultraviolet.  (Recall that the coupling $e^2$ has mass-dimension one.  Hence, without loss of generality, hereafter we set $e^2=1$ and measure all dimensioned quantities with respect to this scale.)

\begin{figure}[t] 
\centerline{
\includegraphics[clip,width=0.8\columnwidth]{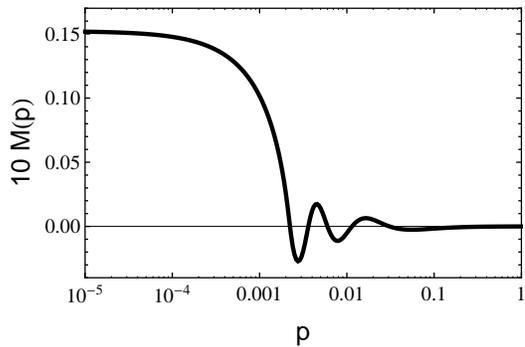}}
\caption{A chiral-limit multi-zero solution for $M(p)$.  Once the mass-scale is set via $e^2=1$, the magnitude of the solution is fixed by the nonlinearity of Eq.\,\protect\eqref{SDE-Mass}.
\label{fig:multinode}}
\end{figure}

In Fig.\,\ref{fig:multinode} we display an example of a multi-zero solution of Eq.\,\protect\eqref{SDE-Mass}.  With the numerical techniques typically employed to solve the gap equation, such solutions are difficult to find and also unstable, in the sense that they are lost when even a very small current-fermion mass ($m_0\ll 1$) is introduced.  Such difficulties are overcome if a more sophisticated numerical algorithm is used; e.g., the homotopy continuation method described and employed in Ref.\,\cite{Wang:2012me}.

Indeed, Ref.\,\cite{Wang:2012me} presents a detailed analysis of the case $m_0\neq 0$.  Little therefore needs to be explained herein.  It suffices to record that, at fixed interaction strength, the number of distinct solutions to the gap equation diminishes rapidly with increasing $m_0$.  Moreover, there is a critical value, $m_0^{\rm cr}=0.0046$ herein, such that Eq.\,\protect\eqref{SDE-Mass} supports only the positive-definite zero-free solution for $m_0>m_0^{\rm cr}$.

\subsection{Dressed vertex in the gap equation}
\subsubsection{Central Ball-Chiu vertex}
It is natural to ask whether the existence of multiple solutions to the gap equation is a peculiar feature of the rainbow truncation.  In partial answer we note that Refs.\,\cite{Fischer:2008sp,Wang:2012me} found this property persisted when the so-called central Ball-Chiu \emph{Ansatz} was employed; viz.,
\begin{equation}
\label{V1BC}
\Gamma_\mu^{1 {\rm BC}}(k,p) = \gamma_\mu\, \frac{1}{2}\left[ \frac{1}{Z(k)} + \frac{1}{Z(p)}\right]\,.
\end{equation}
We will therefore reconsider this case herein, bringing to bear the knowledge we've gained from Sec.\,\ref{Theory-H}.

Using Eq.\,\eqref{V1BC} in Landau gauge, the gap equation produces $Z(p)\equiv 1$ and the following equation for the mass function:
\begin{eqnarray}
%
%
M(p) & = & m_0 + \frac{2 \alpha}{\pi p} \int dk \frac{k  M(k)}{k^2+M^2(k)} \ln \left| \frac{k+p}{k-p}\right|.
\end{eqnarray}
This is not different from Eq.\,\eqref{SDE-Mass} and hence the pattern of behavior described in Secs.\,\ref{sec:linearized}, \ref{sec:nonlinear} is repeated with Eq.\,\eqref{V1BC}.


\subsubsection{Symmetric central vertex}
Now consider a simple extension of Eq.\,\eqref{V1BC}; viz.
\begin{equation}
\label{symmetricV}
\Gamma_\mu(k,p) = \gamma_\mu \, \mathpzc{f}(k^2,p^2) \,,
\end{equation}
where $\mathpzc{f}(k^2,p^2)$ is a symmetric function that approaches unity as either or both of $k^2$, $p^2$ approach infinity.  In this case, again, $Z(p)\equiv 1$ in Landau gauge and the mass function satisfies
\begin{equation}
M(p)= m_0+
 \frac{2 \alpha}{\pi p} \int dk \frac{k M(k) }{k^2+M^2(k)} \mathpzc{f}(k^2,p^2)
\ln \left| \frac{k+p}{k-p}\right|.
\end{equation}
Plainly, therefore, with any function $\mathpzc{f}(k^2,p^2)$ that preserves Assumption\,$\mathbf{G_1}$ in Sec.\,\ref{Theory-H}, the chiral limit gap equation will possess a countable infinity of distinct solutions, distinguished by their number of zeros and their sign and magnitude at $p=0$.  As will become plain in Sec.\,\ref{gauge}, $Z(p)\equiv 1$ is not required to ensure this outcome.  (A further illustration of these general statements is presented in Ref.\,\cite{Zhu:2013zna}.)

\subsubsection{More general \mbox{\emph{Ans\"atze}}}
The class of vertex \emph{Ans\"atze} described by Eq.\,\eqref{symmetricV} is large.  It includes the form in Eq.\,\eqref{V1BC}.  However, it is known that the true dressed-fermion--gauge-boson vertex is more complicated still (especially in the presence of DCSB), involving as many as eleven other Dirac matrix structures \cite{Ball:1980ay,Curtis:1990zs,Kizilersu:2009kg,Chang:2010hb,%
Bashir:2011vg,Bashir:2011dp,Qin:2013mta}.
We have therefore solved the QED3 gap equation using this range of more sophisticated \emph{Ans\"atze} and the numerical algorithms that delivered every solution explained above, irrespective of its complexity (see App.\,\ref{sec:appendix}).  In each of these cases we find the positive definite DCSB solution of the chiral-limit gap equation and its mirror image but no solutions that possess even a single zero.

Naturally, despite the effectiveness of our numerical methods when employed with simple vertex \emph{Ans\"atze}, the explanation might be that they are inadequate to the task when these more complicated and realistic \emph{Ans\"atze} are used.  On the other hand, such \emph{Ans\"atze} produce gap equations that are not of the Hammerstein form.  In fact, with extant vertex \emph{Ans\"atze}, the gap equations are typically elements in the more general class of nonlinear Urysonh integral equations \cite{Krasnoselskii,Polyanin:2008}.  Little is known about the nature of solutions to such equations.  Moreover, the full theory must be even more complicated because, in addition to the explicit dependence on $S(p)$ expressed in \emph{Ans\"atze}, the true dressed vertex will possess an implicit dependence on the dressed-propagator whose nature is impossible to guess.

These observations indicate that the known conditions under which the gap equation can possess multiple solutions are quite strict and, moreover, that they are usually not met when the dressed-fermion--gauge-boson vertex possesses what might justifiably be called a \emph{realistic} form.


As a counterpoint, we have considered a combination of the QCD-based interactions in Refs.\,\cite{Maris:1999nt,Qin:2011dd} and the complete Ball-Chiu \emph{Ansatz}.  In these cases the homotopy continuation algorithm reveals that there are no solutions with zeros unless the interaction strength is inflated to an unrealistically large value; i.e., more than five-times the strength required to explain contemporary experiment \cite{kunlunprivate:2013}.  Owing to the capacities of the homotopy continuation algorithm, which are detailed elsewhere \cite{Wang:2012me}, one can safely conclude that no solution has been overlooked.

Applying this experience to QED3, one can reasonably argue that a QED3 gap equation with the Ball-Chiu vertex possesses no solutions with a zero.  This is because the mass-scale is set by the coupling strength, which therefore does not provide a tool for inflating the interaction strength relative to other scales; and, furthermore, fermion loops provide screening, so that unquenching acts to further suppress the interaction strength.

\section{Gauge covariance}
\label{gauge}
The fermion propagator is gauge covariant but not gauge invariant; and in the preceding discussion we have focused on Landau gauge.  As we noted above, this is because Landau gauge occupies a special place in gauge theories.  In addition to other important properties, such as being a fixed point of the renormalization group and the gauge in which any sensitivity to model-dependent differences between \emph{Ans\"atze} for the fermion-photon vertex are least noticeable, it is also the sole covariant gauge in which the infrared behavior of the fermion propagator is not modified by a non-dynamical gauge-dependent exponential factor whose presence can obscure truly observable features of the theory \cite{Bashir:2008fk,Bashir:2009fv}.  Moreover, as we now explain, by capitalizing on the Landau-Khalatnikov-Fradkin (LKF) transformations \cite{Fradkin:1955jr,LK56,Johnson:1959zz,Zumino:1959wt}, a mechanical realization of the gauge covariance property, one may focus on Landau gauge in Abelian theories without loss of generality.

In configuration space, the fermion propagator in any covariant gauge, $\xi$, is obtained from its Landau gauge ($\xi=0$) form via the following operation:
\begin{equation}
\label{LKFx} S(z;\xi) = S(z;\xi=0)\, {\rm e}^{- \varsigma |z|},\;
\varsigma= \xi \frac{e^2}{8 \pi}= \frac{1}{2} \xi \alpha\, .
\end{equation}
For $\xi>0$, one may readily translate this transformation into momentum space \cite{Bashir:2000ur,Bashir:2002sp,Bashir:2004yt,Bashir:2005wt}:
\begin{subequations}
\label{LKFTs}
\begin{eqnarray}
\nonumber
\sigma_V(p;\xi) & =& \frac{\varsigma}{\pi p^2}\int_0^\infty \!\! dk\, k^2 \sigma_V(k;0) \\
&& \times
\left[\frac{1}{\lambda^ -}+\frac{1}{\lambda^+}+\frac{1}{2kp}
\ln{  {\frac{\lambda^-}{\lambda^+}}  }
\right], \label{gistV}\\
\sigma_S(p;\xi) &=&
\frac{\varsigma}{\pi p}\int_0^\infty \!\! dk\, k\ \sigma_S(k,0)
\left[\frac{1}{\lambda^-}-\frac{1}{\lambda^+}\right]\!, \label{gist}
\end{eqnarray}
\end{subequations}
where $\lambda^\pm = \varsigma^2 + (k\pm p)^2$ and we have used Eq.\,\eqref{Spsigmavs}.  The analogues for $\xi<0$ are given in Ref.\,\cite{Bashir:2009fv}.  Plainly, with a solution for the dressed-quark propagator in hand, obtained using any Landau-gauge \emph{Ansatz} for the dressed-fermion--gauge-boson vertex, one can straightforwardly find the result in another covariant gauge.

\begin{figure}[t] 
\centerline{\includegraphics[clip,width=0.8\linewidth]{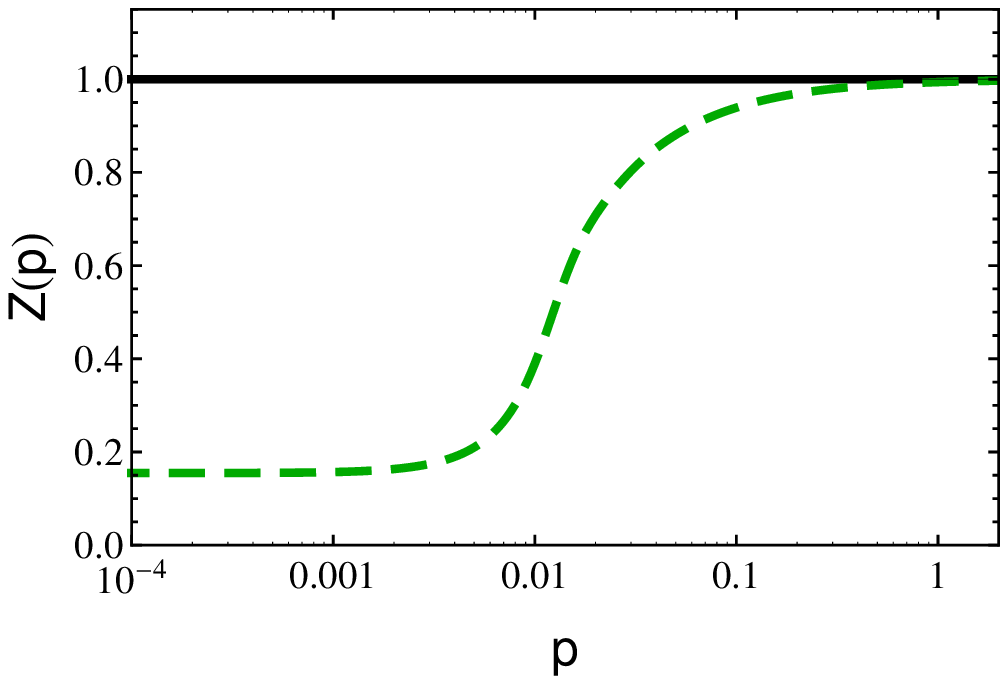}}
\centerline{\includegraphics[clip,width=0.8\linewidth]{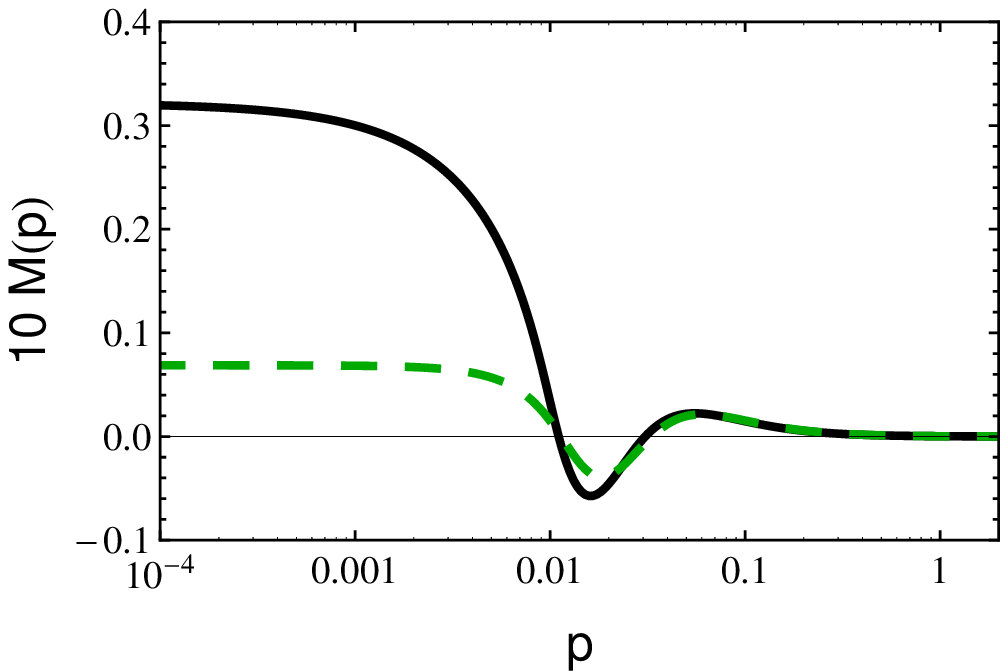}}
\caption{Two-zero solution obtained with Eq.\,\protect\eqref{V1BC}: \emph{upper panel}, $Z(p)$; and \emph{lower panel}, $M(p)$.  Both panels: $\xi=0$, solid curve; and $\xi=0.1$, dashed curve, obtained via Eqs.\,\protect\eqref{LKFTs}. \label{fig:LKF2}}
\end{figure}

We illustrate the result for a two-zero solution in Fig.\,\ref{fig:LKF2}.  It is true in general that the wavefunction renormalization, $Z(p)$, remains zero-free in all covariant gauges and the LKF transformation does not shift the location of any zero in $M(p)$.  It is clear, therefore, that whatever is the qualitative character of the gap equation's Landau gauge solution, it is the same in all covariant gauges.  Namely, if there are multiple solutions for the mass function in Landau gauge, each with a different number of zeros, then these solutions exist in all gauges.  Notably, however, for $\xi>0$ their magnitude is rapidly damped (see Eq.\,\eqref{LKFx} and lower panel of Fig.\,\ref{fig:LKF2}), an effect that will make them difficult to locate numerically if one chooses to solve the gap equation directly in a different gauge.  Of course, the gap equation solution obtained directly in a different gauge would only be meaningful if the gap equation were built with the LKF transform of the original Landau-gauge vertex \emph{Ansatz}.

\section{Epilogue}~\label{final}
Using quenched QED3 as an example, we explored conditions under which a gauge theory's chiral-limit gap equation can exhibit multiple solutions.  We argued that so long as the Landau-gauge gap equation can be rewritten as an Hammerstein integral equation of the first kind, it will possess a countable infinity of simultaneous solutions.  Those solutions will be distinguished, one from another, by the number of zeros they possess and their magnitude and sign at the origin.

Whilst a large class of truncations of gauge theory gap equations possess the Hammerstein property, a far larger class does not.  Membership of the Hammerstein class is not usually determined by the form of the dressed-gauge-boson propagator but, instead, it is decided by the structure of the dressed-fermion--gauge-boson vertex.

We cannot now say anything definite in general about the existence of multiple solutions for members of the non-Hammerstein class of gap equations.  However, by analyzing examples from that class, we were led to conjecture that there is at least a large subclass that do not possess solutions with zeros.  Indeed, it appears to us that the general conditions under which a gauge theory's chiral-limit gap equation can possess solutions with zeros are quite restrictive; and it is therefore probable that when a realistic \emph{Ansatz} is employed for the dressed vertex, the only dynamical chiral symmetry breaking solutions that exist are those with no zeros.

Although we have only considered QED3 explicitly, the mathematical framework we have described applies more broadly.  We therefore judge that our results are equally relevant to existing realistic models for the gap equation in QCD.

\begin{acknowledgments}
We thank V.\,P.~Gusynin, V.\,A.~Miransky, I.\,A.~Shovkovy and K.-l.~Wang for illuminating discussions.  This work received support from CIC (UMSNH) and CONACyT Grant nos.\ 4.10, 4.22, 46614-I and 128534; and
U.\,S.\ Department of Energy, Office of Nuclear Physics, contract no.~DE-AC02-06CH11357.
\end{acknowledgments}

\appendix
\section{Comments on numerical methods}
\label{sec:appendix}
As practitioners will quickly learn, an iterative scheme is unlikely to find a gap equation solution with zeros.  We therefore employed a collocation method.  Namely, upon introducing a quadrature rule, the integral equation is recast as a set of nonlinear algebraic equations.  (This generalises the procedure associated with Eqs.\,\eqref{QuadratureI}, \eqref{Liouville} in our manuscript.)  To solve that system, we employed the secant method and the Broyden strategy, with a bounded trial solution.  (Useful descriptions of solution strategies for nonlinear integral equations may be found in Refs.\,\cite{Bloch:1995dd,AUTARKAW,Broyden:1965}.)  The two approaches produced solutions with zeros in those cases where such solutions were known to exist; e.g., Eqs.\,\eqref{eqgraf}, \eqref{SDE-MassLog} above.

However, when these approaches were adapted to integral equations obtained with the vertex \emph{Ans\"atze} in Refs.\,\cite{Ball:1980ay,Curtis:1990zs,Kizilersu:2009kg,Chang:2010hb,%
Bashir:2011vg,Bashir:2011dp,Qin:2013mta}, which are not Hammerstein equations, only solutions without zeros were obtained.

As a step toward verifying this outcome, we then expanded the arguments of the integral equation in terms of a complete set of functions and solved for the expansion coefficients.  The solutions without zeros were recovered but no new solutions were found.  In this approach we focused on the ``sinc'' and ``little sinc'' collocation methods \cite{stenger,0305-4470-39-22-L01,1751-8121-40-43-013}.


\end{document}